\documentclass[12pt]{article}

\usepackage{amsmath,amssymb}
  \usepackage{authblk}
\usepackage{hyperref}  
\usepackage{cite}

\usepackage{enumitem}  
 
\usepackage[usenames, dvipsnames]{color} 
\usepackage{graphicx}     
\usepackage{subcaption}
\usepackage[utf8]{inputenc}
\usepackage{url}
 \usepackage{xcolor}

\def\beq{\begin{equation}}
\def\eeq{\end{equation}}

\def\d{\delta}

\def\S{\Sigma}

\begin{document}

\thispagestyle{empty}

\centerline{{\Large \bf   Spacetime Equilibrium at Negative Temperature }} 
\vskip.5truecm
\centerline{{\Large \bf    and the Attraction of Gravity}}
\vskip.95truecm

\begin{center}
{\large Ted Jacobson$^{1 \, *}$ and Manus Visser$^{2 \, \dagger}$}\\
\vskip .75truecm
$^1${\it Maryland Center for Fundamental Physics,  University of Maryland, College Park, MD 20742, USA}\\[1mm]
$^2${\it Institute for Theoretical Physics,  University of Amsterdam, \\ 1090 GL Amsterdam, The Netherlands}\\
\vskip .75truecm
{$^{*}$\tt jacobson@umd.edu}$\quad$ {$^{\dagger}$\tt m.r.visser@uva.nl}

\vskip 1cm

\end{center}
\vskip 1.5cm

\begin{abstract}

We derive the  Einstein equation   from  the condition that every small causal diamond is a   variation of a flat empty diamond with the same free conformal energy, as would be expected for a near-equilibrium state.  The attractiveness of gravity hinges on the negativity of   the absolute temperature of these diamonds, a property we infer from the generalized entropy. 
 
\end{abstract}

\thispagestyle{empty}

\vskip 5cm

\begin{center}
{ \it Essay written for the Gravity Research Foundation 2019 Awards \\for Essays on Gravitation; submitted March 31, 2019.}
\end{center}

\newpage 
\setcounter{page}{1}

The discovery of the Unruh effect \cite{Unruh:1976db} revealed that the distinction between vacuum fluctuations and thermal fluctuations is not as great as previously thought \cite{Sciama:1981hr}.
Indeed, the simplest and most general statement of this relation is that, for all observables 
localized in a Rindler wedge, the Minkowski vacuum of a relativistic quantum field is a thermal state
with respect to the Lorentz boost Hamiltonian. Since every point in any spacetime has an
approximately Minkowskian neighborhood, one is led to the idea that spacetime can be 
viewed as a medium, everywhere near local thermodynamic equilibrium, somewhat like a fluid with local temperature,
density, pressure, etc. \cite{Jacobson:1995ab}.

The entanglement entropy of a quantum field vacuum across a Rindler horizon is
UV divergent, and scales with the horizon area $A$ \cite{Sorkin:2014kta,Bombelli:1986rw,Srednicki:1993im}. 
In quantum gravity, the UV divergent area law for entropy 
is presumably replaced by the finite, Bekenstein-Hawking area law, $S_{\rm BH} = A/4\hbar G$ \cite{Bekenstein:1973ur,Hawking:1974sw}, thus establishing a link between 
gravitation and thermodynamics of the vacuum. In the context of black hole thermodynamics,
Bekenstein introduced the ``generalized entropy", 
$S_{\rm gen} : = S_{\rm BH} + S_{\rm m}$, where $S_{\rm m}$ is the ordinary matter 
entropy outside the horizon, and he argued that  
$S_{\rm gen}$ should satisfy the ``generalized second law" (GSL)
of thermodynamics, thanks to the Einstein equation. 
All evidence suggests that he was right, and that the  GSL holds not
only globally for black holes, but also locally for Rindler horizons\cite{Wall:2010cj,Wall:2011hj}.

In Ref.~\cite{Jacobson:2015hqa}, reversing the logic,
the Einstein equation was derived from the equilibrium assumption
that the generalized entropy of small 
causal diamonds is stationary at fixed spatial volume. 
Sorely lacking, however, was a prior rationale for holding fixed the volume.
In this essay (see also \cite{Jacobson:2018ahi}) we reformulate the derivation of \cite{Jacobson:2015hqa},
replacing the stationarity of entropy at fixed volume by the 
stationarity of a free energy. 
The volume appears in the 
free energy, playing the role of the gravitational energy. 
That the volume should play this role can
be derived from general relativity, and is related \cite{Jacobson:2018ahi} to the observation of York \cite{York:1972sj} that the 
Hamiltonian of general relativity in the extrinsic curvature time gauge is proportional to the 
spatial volume of constant mean curvature slices. Here, however, since we aim to {\it derive}
the Einstein equation, we cannot use results from general relativity.  Instead, 
we infer from  diffeomorphism invariance the need for a volume term in the free energy. 
Ultimately, perhaps a microscopic interpretation of the volume as some kind of energy can be found, 
in the same way that the area represents entanglement entropy.
 
Our key postulate is that any small causal diamond in any spacetime is ``close" to being 
flat (Minkowski), in the sense that, to first order in 
metric and matter variations, it is the variation of a flat reference causal diamond with the same free energy in the canonical ensemble. 
To qualify as ``small" for this purpose, a diamond should be much smaller than both the shortest local curvature scale and the scale of any quantum field excitations present.  
 Since all metrics are   flat to first order around any point, a small diamond can be regarded as a slight deformation of a flat diamond, 
i.e.\ the intersection of the future of one point with the past of another 
in Minkowski spacetime.  Such diamonds admit
  a conformal isometry generated by a conformal Killing vector $\zeta$   \cite{Hislop:1981uh,Faulkner:2013ica}, 
satisfying $\mathcal L_\zeta g_{ab} \propto g_{ab}$,  
which is null at the past and future null boundaries of the     diamond, so those boundaries are conformal Killing horizons.  The horizon surface gravity~$\kappa$
(defined through $\nabla_a \zeta^2 =- 2 \kappa \zeta_a$ \cite{Jacobson:1993pf}) is constant on the edge of the diamond by spherical symmetry, as well as along each generator of the null boundaries, as for the horizon of a stationary black hole.
We normalize the conformal Killing vector below 
such that $\kappa=1$.

We define the Helmholtz-like free (conformal) energy of flat  diamonds as
 \begin{equation} \label{freeenergy}
 F=     H_\zeta   - T S_{\rm UV}   . 
 \end{equation}
 Here,  $  H_\zeta $ is the Hamiltonian generating evolution along the flow of the conformal Killing field $\zeta$, $T$ is the temperature of the diamond, and $S_{\rm UV}$ is the entanglement entropy of the diamond associated with the UV    degrees of freedom.  One might have thought that a true Killing vector is needed for gravitational thermal equilibrium, however we find that for causal diamonds a conformal Killing vector suffices. 
We assume there exists an UV-cutoff in quantum spacetime which renders the entanglement entropy finite \cite{Frolov:1993ym,Jacobson:1994iw,Jacobson:2012yt}, and proportional to the area,
\begin{equation} \label{tempent}
S_{\rm UV} = \eta   A \, , 
 \end{equation} 
where $\eta$ is a universal positive constant of dimension $[\text{length}]^{2-d}$. 
As to the temperature, one might think it should be  
the Unruh temperature associated with the conformal Killing horizon, $T_{\rm U} = \hbar/2\pi$,
since, for conformal matter on a background flat diamond, 
the variation of the matter Hamiltonian away from the conformal vacuum 
is equal to $T_{\rm U}$ times the variation of the matter entropy, 
$\delta H^{\rm m}_\zeta = T_{\rm U}\d S^{\rm m}$. 
(This is the conformal generalization of the 
Unruh effect \cite{Hislop:1981uh}.) 
When inserted into the variation (at fixed $T$) of the free energy~\eqref{freeenergy}
this yields 
\beq
\d F\supset   T_{\rm U}\d S^{\rm m} - T \d S_{\rm UV}.
\eeq
The two terms on the right  should combine to form $-T\d S_{\rm gen}$, but this 
happens only if 
\beq \label{temp}
T=-T_{\rm U}=-\hbar/2\pi.
\eeq
Thus, quantum field thermodynamics   in a {\it fixed}   diamond background is quite different from the self-gravitating case. In the former the temperature of the vacuum is positive, whereas in {\it gravitational} thermodynamics, as we see here, the temperature of a causal diamond is negative. 
 
The  conformal Killing energy   $H_\zeta  $ has   contributions from  the metric and from matter fields.   We work in the semiclassical regime, i.e. we consider   \emph{quantum} matter fields on a \emph{classical} background spacetime.  For the stationarity of free energy, we only need to know the {\it variation} of the conformal Killing energy,  denoted by: $\delta H_\zeta = \delta \langle H_\zeta^{\rm m} \rangle + \delta H_\zeta^{\rm g}$. The variations we consider are arbitrary variations of the dynamical fields away from the flat   diamond to nearby states.   The variation of the expectation value of the matter Hamiltonian is given by an integral over the maximal slice $\S$ (which is a spherical ball) of the reference diamond,\footnote{For a traceless and divergencefree stress tensor   the integral would be independent of the slice.}  
\begin{equation} \label{matterHam}
\delta \langle H_\zeta^{\rm m} \rangle = \int_\Sigma  \delta  \langle  {T_{a}}^b \rangle \zeta^a u_b dV   .
\end{equation}
We   take the reference configuration to be one with vanishing stress  tensor;   in effect, $ \delta  \langle  {T_{a}}^b \rangle =      \langle  {T_{a}}^b \rangle$. Note that, since we have not  converted the matter stress-energy into an equivalent entanglement entropy,  non-conformal invariant matter presents no extra complications   in our derivation, unlike in \cite{Jacobson:2015hqa} where an extension of the first law of entanglement entropy was required.

The gravitational contribution $ \delta   H_\zeta^{\rm g}$  to the Hamiltonian variation  can be inferred, without assuming the Einstein-Hilbert action, from the requirement of diffeomorphism invariance. 
Consider a variation induced by a diffeomorphism, denoted by $\hat \delta$. 
 In that case $\hat \delta \langle H_\zeta^{\rm m} \rangle $ is zero,  since   the background value is taken to vanish. 
 Stationarity of the free conformal energy \eqref{freeenergy}  at fixed temperature then implies: $\hat \delta H_\zeta^{\rm g} = T\eta \hat\delta  A$, where we used \eqref{tempent}.  In Section~3.3.2 of \cite{Jacobson:2018ahi} we showed that the diffeo-induced area variation is equal to: $\hat \delta A = k \hat \delta  V$, where $\hat \delta V$  is the variation of the   volume of the maximal slice in the original diamond 
 and $k$ is the trace of the extrinsic curvature of the edge, as embedded in the maximal slice. 
 The minimal choice for $\delta H_\zeta^{\rm g}$ consistent with diffeomorphism invariance is thus given just by the volume variation. 
 Hence, we postulate that the gravitational Hamiltonian variation is equal to  
 \begin{equation} \label{gravHam}
 \delta   H_\zeta^{\rm g}    = T \eta \,  k \,  \delta V  . 
 \end{equation}
The   variation of the free energy \eqref{freeenergy} at fixed temperature is therefore    
\begin{equation} \label{FEvar}
\delta F =  \delta \langle H_\zeta^{\rm m} \rangle - T \eta \left (  \delta A - k \delta V \right).
\end{equation} 
Note that the definitions \eqref{tempent}, \eqref{temp}, \eqref{matterHam} and \eqref{gravHam}  
of the terms in the free energy (variation) apply in principle to flat diamonds of \emph{any} size.

Next, we evaluate $\delta F$  for \emph{small} diamonds.  
For such diamonds, to leading order  in $\ell/L_{\rm excitation}$,      
  the matter contribution to the Hamiltonian variation  is    \cite{Jacobson:2015hqa, Jacobson:2018ahi}
 \begin{equation} \label{matterHamsmall}
 \delta \langle H_\zeta^{\rm m} \rangle = \frac{ \Omega_{d-2} \ell^d }{d^2-1}   \langle {T_{ab}} \rangle u^a u^b \, , 
 \end{equation}
 where $\Omega_{d-2}$ is the area of a unit $(d-2)$-sphere, $d$ is the spacetime dimension and $ \langle {T_{ab}} \rangle u^a u^b$ is  {constant to leading order on the maximal slice.\footnote{The fraction on the right-hand   side  of   \eqref{matterHamsmall} originates from the integral of the norm of the conformal Killing vector over de volume of the ball: $ \int_\Sigma |\zeta| dV$. This integral is   known as the ``thermodynamic volume'' in the case where $\zeta^a$ is a true Killing vector \cite{Kastor:2009wy}.} 
  Meanwhile, to    lowest order in $\ell / L_{\rm curvature}$,
 we have  
 \begin{equation} \label{AandVsmall}
 \delta A - k \delta V = - \frac{\Omega_{d-2}\ell^d}{d^2-1}  G_{ab}   u^a u^b \, ,
 \end{equation}
 where the Einstein curvature tensor $G_{ab}$ is evaluated at the center of $\Sigma$.   This purely geometric result was obtained in Ref.  \cite{Jacobson:2015hqa} using a Riemann normal coordinate expansion. 
  Finally, inserting   \eqref{matterHamsmall} and \eqref{AandVsmall}   into the expression \eqref{FEvar} for the free energy variation yields
 \begin{equation} \label{freeenergyvar2}
 \delta F 
 = \frac{\Omega_{d-2} \ell^d}{d^2 -1}  u^a u^b\left (    \langle T_{ab} \rangle + T \eta   \,   G_{ab}   \right)  . 
 \end{equation}
 Note   that the \emph{same} fraction appears  in the combination of variations \eqref{AandVsmall} as in the matter Hamiltonian variation \eqref{matterHamsmall}, which is crucial for the agreement of the entropy area-density $\eta$  with the Bekenstein-Hawking value to follow from the   Einstein equation \eqref{Eineq}.
 
The requirement  $\delta F = 0$    for all timelike unit normals $u^a$ and at every point  in spacetime now  implies  the equation
\begin{equation} \label{Eineq}
G_{ab}   = - \frac{1}{T \eta}  \langle T_{ab} \rangle   .
\end{equation}
Recall that we previously inferred the diamond temperature    $T =-\hbar/2\pi $. Thus,  equation \eqref{Eineq} is  the semiclassical Einstein equation   provided we make the    identification 
$ 
 \eta =1/4  \hbar G 
$
(where $G$ is Newton's constant),
which agrees with the Bekenstein-Hawking  entropy area-density. 
 Note that the emergent gravity is attractive since the temperature is negative, and would have been repulsive had the temperature been positive! A cosmological constant is of course permitted as a    piece of the stress tensor equal to a constant times  the metric.

  The negative temperature  is a surprising feature  of our derivation. 
 That it must be negative is already evident from classical Einstein gravity, since 
the addition of energy to a diamond results in the {\it decrease} of its Bekenstein-Hawking entropy at fixed volume   \cite{Jacobson:2015hqa,Jacobson:2018ahi}. 
Negative temperature typically requires of a system that i) the energy spectrum is bounded from above, and ii) the Hilbert space is finite-dimensional. 
As argued  by Klemm and Vanzo \cite{Klemm:2004mb} for the de Sitter   static patch,    causal diamonds  indeed satisfy these properties: i) there is        an upper bound on the energy, equal to the    mass of the largest black hole that fits inside a diamond  given its bounding area, and ii) the holographic principle implies  that the entropy of a causal diamond is bounded by the Bekenstein-Hawking entropy associated to the  area of the edge. Thus, despite the positive value of the Unruh temperature, we must  conclude that the temperature of a self-gravitating causal diamond is negative.\\ \\

\noindent TJ was supported in part by NSF grants PHY-1407744 and PHY-1708139. MV acknowledges support from the Spinoza Grant of the Dutch Science Organisation (NWO), and from the Delta ITP consortium, a special program of the NWO. We thank the organizers of the ``Quantum gravity and quantum information'' CERN Theory Institute, where this essay was conceived.

\newpage

\bibliographystyle{JHEP}
\bibliography{equilibrium}

\providecommand{\href}[2]{#2}\begingroup\raggedright\begin{thebibliography}{10}

\bibitem{Unruh:1976db}
W.~G. Unruh, \emph{{Notes on black hole evaporation}},
  \href{https://doi.org/10.1103/PhysRevD.14.870}{\emph{Phys. Rev.} {\bfseries
  D14} (1976) 870}.

\bibitem{Sciama:1981hr}
D.~W. Sciama, P.~Candelas and D.~Deutsch, \emph{{Quantum Field Theory, Horizons
  and Thermodynamics}},
  \href{https://doi.org/10.1080/00018738100101457}{\emph{Adv. Phys.} {\bfseries
  30} (1981) 327}.

\bibitem{Jacobson:1995ab}
T.~Jacobson, \emph{{Thermodynamics of space-time: The Einstein equation of
  state}}, \href{https://doi.org/10.1103/PhysRevLett.75.1260}{\emph{Phys. Rev.
  Lett.} {\bfseries 75} (1995) 1260}
  [\href{https://arxiv.org/abs/gr-qc/9504004}{{\ttfamily gr-qc/9504004}}].

\bibitem{Sorkin:2014kta}
R.~D. Sorkin, \emph{{1983 paper on entanglement entropy: ``On the Entropy of
  the Vacuum outside a Horizon"}},  in \emph{{Proceedings, 10th International
  Conference on General Relativity and Gravitation: Padua, Italy, July 4-9,
  1983}}, vol.~2, pp.~734--736, 1984,
  \href{https://arxiv.org/abs/1402.3589}{{\ttfamily 1402.3589}}.

\bibitem{Bombelli:1986rw}
L.~Bombelli, R.~K. Koul, J.~Lee and R.~D. Sorkin, \emph{{A Quantum Source of
  Entropy for Black Holes}},
  \href{https://doi.org/10.1103/PhysRevD.34.373}{\emph{Phys. Rev.} {\bfseries
  D34} (1986) 373}.

\bibitem{Srednicki:1993im}
M.~Srednicki, \emph{{Entropy and area}},
  \href{https://doi.org/10.1103/PhysRevLett.71.666}{\emph{Phys. Rev. Lett.}
  {\bfseries 71} (1993) 666}
  [\href{https://arxiv.org/abs/hep-th/9303048}{{\ttfamily hep-th/9303048}}].

\bibitem{Bekenstein:1973ur}
J.~D. Bekenstein, \emph{{Black holes and entropy}},
  \href{https://doi.org/10.1103/PhysRevD.7.2333}{\emph{Phys. Rev.} {\bfseries
  D7} (1973) 2333}.

\bibitem{Hawking:1974sw}
S.~W. Hawking, \emph{{Particle Creation by Black Holes}},
  \href{https://doi.org/10.1007/BF02345020, 10.1007/BF01608497}{\emph{Commun.
  Math. Phys.} {\bfseries 43} (1975) 199}.

\bibitem{Wall:2010cj}
A.~C. Wall, \emph{{A Proof of the generalized second law for rapidly-evolving
  Rindler horizons}},
  \href{https://doi.org/10.1103/PhysRevD.82.124019}{\emph{Phys. Rev.}
  {\bfseries D82} (2010) 124019}
  [\href{https://arxiv.org/abs/1007.1493}{{\ttfamily 1007.1493}}].

\bibitem{Wall:2011hj}
A.~C. Wall, \emph{{A proof of the generalized second law for rapidly changing
  fields and arbitrary horizon slices}},
  \href{https://doi.org/10.1103/PhysRevD.87.069904,
  10.1103/PhysRevD.85.104049}{\emph{Phys. Rev.} {\bfseries D85} (2012) 104049}
  [\href{https://arxiv.org/abs/1105.3445}{{\ttfamily 1105.3445}}].

\bibitem{Jacobson:2015hqa}
T.~Jacobson, \emph{{Entanglement Equilibrium and the Einstein Equation}},
  \href{https://doi.org/10.1103/PhysRevLett.116.201101}{\emph{Phys. Rev. Lett.}
  {\bfseries 116} (2016) 201101}
  [\href{https://arxiv.org/abs/1505.04753}{{\ttfamily 1505.04753}}].

\bibitem{Jacobson:2018ahi}
T.~Jacobson and M.~Visser, \emph{{Gravitational Thermodynamics of Causal
  Diamonds in (A)dS}},  \href{https://arxiv.org/abs/1812.01596}{{\ttfamily
  1812.01596}}.

\bibitem{York:1972sj}
J.~W. York, Jr., \emph{{Role of conformal three geometry in the dynamics of
  gravitation}}, \href{https://doi.org/10.1103/PhysRevLett.28.1082}{\emph{Phys.
  Rev. Lett.} {\bfseries 28} (1972) 1082}.

\bibitem{Hislop:1981uh}
P.~D. Hislop and R.~Longo, \emph{{Modular Structure of the Local Algebras
  Associated With the Free Massless Scalar Field Theory}},
  \href{https://doi.org/10.1007/BF01208372}{\emph{Commun. Math. Phys.}
  {\bfseries 84} (1982) 71}.

\bibitem{Faulkner:2013ica}
T.~Faulkner, M.~Guica, T.~Hartman, R.~C. Myers and M.~Van~Raamsdonk,
  \emph{{Gravitation from Entanglement in Holographic CFTs}},
  \href{https://doi.org/10.1007/JHEP03(2014)051}{\emph{JHEP} {\bfseries 03}
  (2014) 051} [\href{https://arxiv.org/abs/1312.7856}{{\ttfamily 1312.7856}}].

\bibitem{Jacobson:1993pf}
T.~Jacobson and G.~Kang, \emph{{Conformal invariance of black hole
  temperature}},
  \href{https://doi.org/10.1088/0264-9381/10/11/002}{\emph{Class. Quant. Grav.}
  {\bfseries 10} (1993) L201}
  [\href{https://arxiv.org/abs/gr-qc/9307002}{{\ttfamily gr-qc/9307002}}].

\bibitem{Frolov:1993ym}
V.~P. Frolov and I.~Novikov, \emph{{Dynamical origin of the entropy of a black
  hole}}, \href{https://doi.org/10.1103/PhysRevD.48.4545}{\emph{Phys. Rev.}
  {\bfseries D48} (1993) 4545}
  [\href{https://arxiv.org/abs/gr-qc/9309001}{{\ttfamily gr-qc/9309001}}].

\bibitem{Jacobson:1994iw}
T.~Jacobson, \emph{{Black hole entropy and induced gravity}},
  \href{https://arxiv.org/abs/gr-qc/9404039}{{\ttfamily gr-qc/9404039}}.

\bibitem{Jacobson:2012yt}
T.~Jacobson, \emph{{Gravitation and vacuum entanglement entropy}},
  \href{https://doi.org/10.1142/S0218271812420060}{\emph{Int. J. Mod. Phys.}
  {\bfseries D21} (2012) 1242006}
  [\href{https://arxiv.org/abs/1204.6349}{{\ttfamily 1204.6349}}].

\bibitem{Kastor:2009wy}
D.~Kastor, S.~Ray and J.~Traschen, \emph{{Enthalpy and the Mechanics of AdS
  Black Holes}},
  \href{https://doi.org/10.1088/0264-9381/26/19/195011}{\emph{Class. Quant.
  Grav.} {\bfseries 26} (2009) 195011}
  [\href{https://arxiv.org/abs/0904.2765}{{\ttfamily 0904.2765}}].

\bibitem{Klemm:2004mb}
D.~Klemm and L.~Vanzo, \emph{{Aspects of quantum gravity in de Sitter spaces}},
  \href{https://doi.org/10.1088/1475-7516/2004/11/006}{\emph{JCAP} {\bfseries
  0411} (2004) 006} [\href{https://arxiv.org/abs/hep-th/0407255}{{\ttfamily
  hep-th/0407255}}].

\end{thebibliography}\endgroup

\end{document}